%% file: ICHEP2020_Pachmayer.tex
\documentclass[a4paper,11pt]{article}
\usepackage{pos}
\usepackage{graphicx}
\usepackage{floatflt}
\usepackage{float}
\usepackage{wrapfig}
\usepackage{setspace}
\usepackage{lineno}
\include{commands}

\title{Heavy Ions \\
Experimental Overview}
\ShortTitle{Heavy Ions, Experimental Overview}

\author*[a]{Yvonne Pachmayer}

\affiliation[a]{Physikalisches Institut der Universität Heidelberg, \\ Im Neuenheimer Feld 226, D-69120 Heidelberg, Germany}

\emailAdd{pachmay@physi.uni-heidelberg.de}

\abstract{

This article gives an overview of recent highlights from experimental measurements of heavy-ion collisions at ultra-relativistic energies: Measurements of electroweak probes constrain both the initial collision geometry and the nuclear parton distribution functions. Results from soft particle production show that the abundance of light-flavour hadrons from pions up to hypertriton and $^4$He can be described by a universal temperature and that these participate in the collective motion of the system. There are hints of these effects also in small systems, which will be further investigated in future to understand the underlying mechanisms. Studies of hard probes, such as heavy quarks and jets show that parton energy loss plays an important role in heavy-ion collisions. Differential measurements of J/$\psi$ mesons elucidate their production mechanism, i.e.\ regeneration, and give evidence for deconfinement in Pb--Pb collisions at LHC full energy. The large data samples at the LHC enable studies of rare probes such as $\chi_{\rm c1}$(3872) and top--anti-top production. Further, measurements of antinuclei cross sections can provide input for dark matter searches.

}

\FullConference{%
  40th International Conference on High Energy physics - ICHEP2020\\
  July 28 - August 6, 2020\\
  Prague, Czech Republic (virtual meeting)
}

\begin{document}
\maketitle
\vspace{-0.2cm}
\section{Introduction}
\vspace{-0.3cm}
One of the main goals in modern heavy-ion physics is the study of QCD matter at extreme conditions of high temperature and/or high net-baryon number density ($\epsilon \approx$~1~GeV/fm$^3$). Under these conditions, which can be generated in ultra-relativistic heavy-ion collisions at laboratories using accelerators such as the Large Hadron Collider (LHC) at CERN, a colour-deconfined state of strongly-interacting matter, the quark--gluon plasma (QGP)~\cite{Shuryak:1977ut} is created. At RHIC and LHC energies, QCD matter is studied at nearly vanishing net-baryon density (baryochemical potential $\mu_{\rm B}  \approx 0$), for which lattice QCD calculations predict a smooth crossover between hadronic matter and a QGP, where chiral symmetry is restored, at a pseudo-critical temperature $T_{\rm C} \approx 156.5 \pm 1.5$~MeV~\cite{Bazavov:2018mes}.  Furthermore, the QGP is assumed to have existed up to a few microseconds after the Big Bang~\cite{Boyanovsky:2006bf} and might exist in the core of neutron stars~\cite{Alford:2013pma}. Therefore, studies of the QGP will also provide input for e.g.\ theoretical models of the early universe and thus address some of the fundamental questions of mankind. \\
An ultra-relativistic heavy-ion collision can be subdivided into the following stages: i) initial collisions, which occur during the passing of the Lorentz-contracted nuclei; ii)  creation and thermalisation of the QGP during 0.1--1~fm/$c$; iii) expansion, collective motion, and cooling of the QGP; iv) hadronisation once the temperature drops below $T_{\rm C}$; v) after 5--10 fm/$c$~\cite{Braun-Munzinger:2015hba} chemical freeze-out (inelastic collisions stop and hadron yields become fixed)  and vi) kinetic-freeze out occur (elastic collisions stop and momenta no longer change). The final-state particles or their decay particles can then be measured in a detector to infer the characteristics of the QGP. It is not possible to detect the QGP directly, because neither quarks nor gluons are colour-neutral objects. However, different experimental measurements are more sensitive to the various stages of the collision, and by comparing the results with theoretical calculations, various properties of the QGP can be determined. \\
In order to draw firm conclusions on effects induced by the hot medium (‘final-state effects’), the same measurements are performed in pp collisions as a baseline and in proton--nucleus (pA) collisions to study initial-state effects in cold nuclear matter (CNM), such as the modifications of the parton distribution functions (PDF) in the nucleus with respect to that in the free proton. However, it has turned out that these measurements have provided in the last years far more interesting results than being a pure reference (see below). \\
The effects of the hot medium can be quantified using the nuclear modification factor $R_{\rm AA}$, which is defined as the ratio of the $p_{\rm T}$ distributions measured in AA collisions with respect to that in pp collisions:
$R_{\rm AA} = \frac{1}{\langle N_{\rm coll} \rangle}\frac{{\rm d}N_{\rm AA}/{\rm d}p_{\rm T}}{{\rm d}N_{\rm pp}/{\rm d}p_{\rm T}}$   ,
where ${\rm d}N_{\rm AA}/{\rm d}p_{\rm T}$ and ${\rm d}N_{\rm pp}/{\rm d}p_{\rm T}$ are the $p_{\rm T}$-differential yields of a given particle species in AA and pp collisions, respectively, and $\langle N_{\rm coll} \rangle$ is the average number of binary nucleon--nucleon collisions in the overlap region of the colliding nuclei. 
If in-medium effects are absent and cold nuclear matter effects are small, the ratio $R_{\rm AA}$ is unity in the region where particle production from hard scattering processes dominates ($p_{\rm T} \gtrsim$~2~GeV/$c$). 

\vspace{-0.2cm}
\section{Initial State}
\begin{figure}[tbp]
\includegraphics[width=0.42\textwidth]{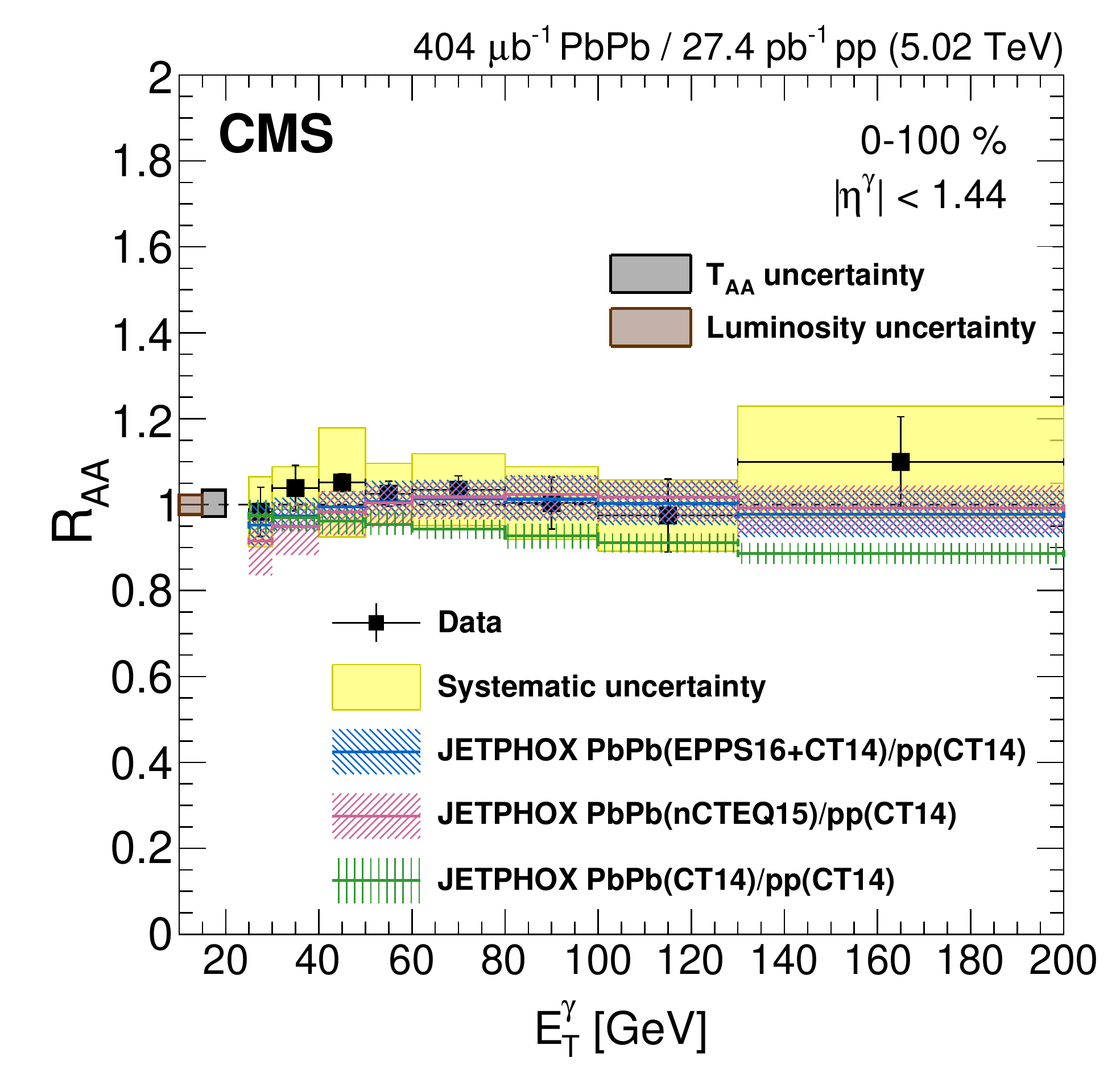}
\hspace{1.0cm}
\includegraphics[width=0.46\textwidth]{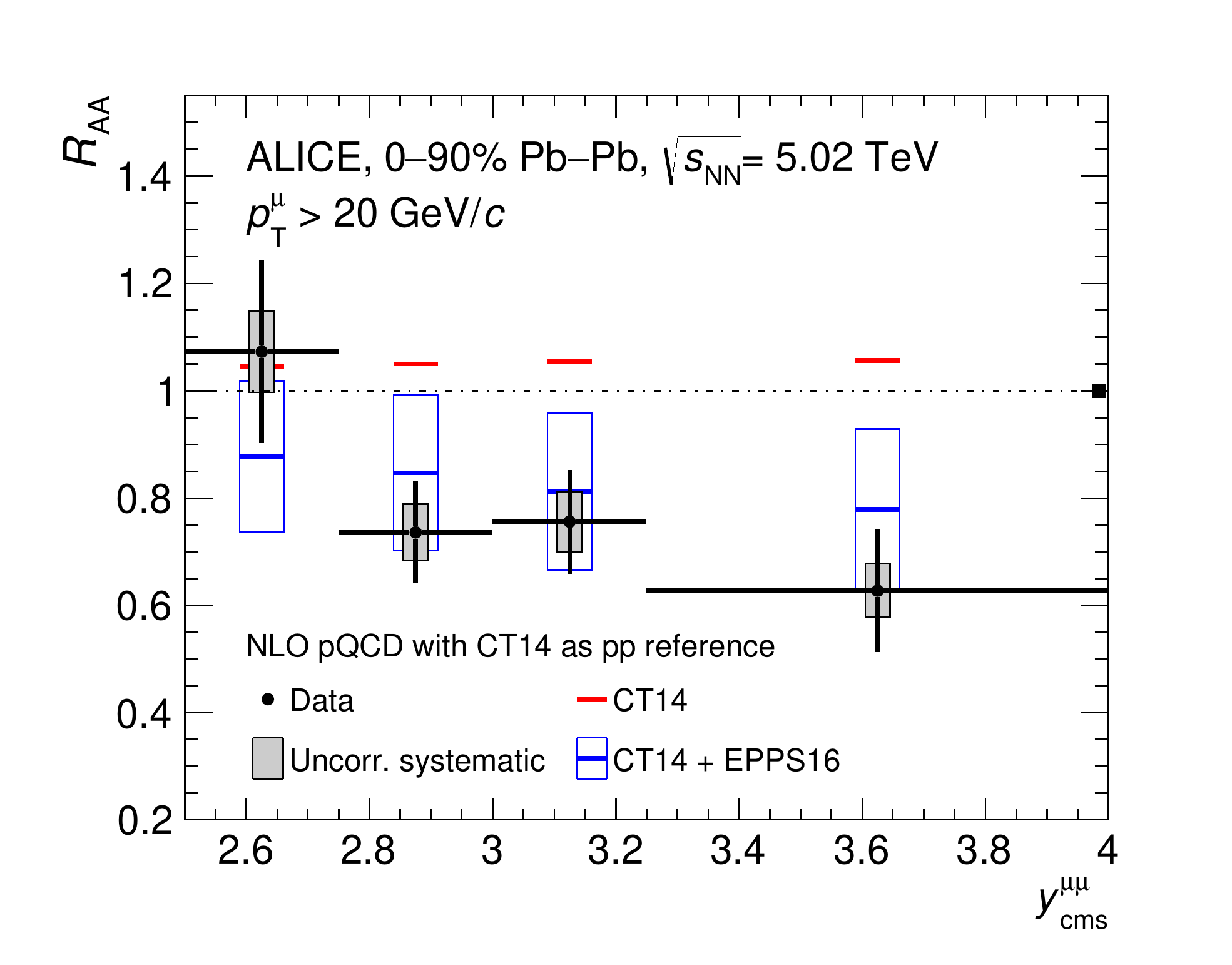}
 \caption{Nuclear modification factor of (left) isolated photons as a function of photon energy~\cite{Sirunyan:2020ycu} and (right) Z bosons vs rapidity~\cite{Acharya:2020puh} in Pb--Pb collisions. \vspace{-0.2cm}} \label{Fig::photons}
\end{figure}

\vspace{-0.2cm}
Photons with high transverse momentum as well as W and Z bosons (due to their large mass, they are produced before the QGP is created), are penetrating probes and provide information about the initial state and the nuclear parton distribution functions. Due to the high collision energy at the LHC, these are copiously produced. The $R_{\rm AA}$ of isolated photons in AA collisions is consistent with unity~\cite{Sirunyan:2020ycu} (see Fig.~\ref{Fig::photons}), since they do not carry colour charge and thus do not interact strongly. The data are better described by calculations with EPPS16 and nCTEQ15 nuclear PDFs than by calculations without nuclear modifications. Measurements by the ATLAS collaboration in p--Pb collisions at forward, mid- and backward rapidity are described by calculations that include modestly modified nPDFs~\cite{Aaboud:2019tab}. The measurement of Z bosons by the ALICE collaboration in Pb--Pb collisions shown in Fig.~\ref{Fig::photons} exhibits a clear deviation from the expectation using free-nucleon PDFs as a function of rapidity, yielding a 3.4$\sigma$ effect integrated over rapidity~\cite{Acharya:2020puh}. Measurements of Z boson production in p--Pb collisions at forward and backward rapidity by the LHCb collaboration agree with calculations considering PDF sets with and without nuclear modification within current uncertainties~\cite{LHCb-CONF-2019-003}. These measurements can be used to better constrain initial state effects and nPDF global fits. 

\vspace{-0.2cm}
\section{Soft Probes}
\vspace{-0.3cm}

The particle yields of light-flavour hadrons and nuclei from pions up to (anti-)hypertriton and (anti-)$^4$He, as measured by the ALICE collaboration in central Pb--Pb collisions at $\sqrt{s_{\rm NN}} =$ 2.76~TeV, span nine orders of magnitude and are well described by a thermal equilibrium model~\cite{Andronic:2017pug} underlining the thermal nature of particle production in ultra-relativistic heavy-ion collisions. The grand canonical thermal fit describes the particle abundances with a universal temperature $T_{\rm chem} = 156 \pm 1.5$~MeV~\cite{Andronic:2017pug}, which coincides with the pseudo-critical temperature of the chiral crossover transition from lattice QCD calculations (see above). 

\begin{wrapfigure}{l}{0.65\textwidth}
\vspace{-0.8cm}
\begin{flushleft}
\includegraphics[width=0.65\textwidth]{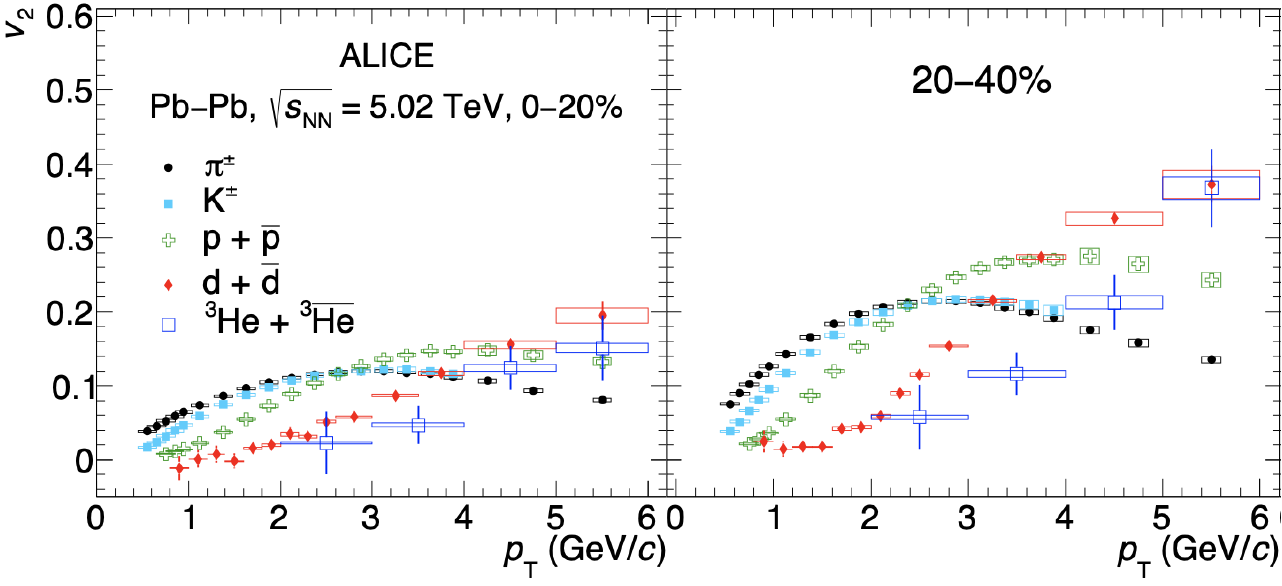}
 \caption{Elliptic flow as a function of $p_{\rm T}$ for various particle species~\cite{Acharya:2019ttn,Acharya:2020lus}.} \label{Fig::flow1}
\end{flushleft}
\vspace{-0.6cm}
\end{wrapfigure}

The azimuthal distributions of particle production with respect to the measured symmetry plane orientations can be expanded in terms of Fourier coefficients $v_{\rm n}$. The `elliptic flow' coefficient $v_{2}$ is shown in Fig.~\ref{Fig::flow1} for different particle species in central and semi-central collisions. For all shown particle species large anisotropies due to collective effects are visible. For semi-central collisions the coefficients are larger due to the initial conditions of the collision. Moreover, at low momenta a mass ordering effect is visible characteristic for hydrodynamic collective expansion. Both deuterons and $^3$He are loosely bound composite objects and the data can also be used to understand the hadronisation mechanism. The $^3$He data are well described by a model that includes coalescence of nucleons, a hydrodynamic evolution of the fireball and a hadronic afterburner~\cite{Zhao:2018lyf}. The calculations also reveal that the produced matter has a small shear viscosity to entropy density ratio, i.e.\ the created fireball is an almost perfect liquid. \\

\begin{figure}[btp]
\includegraphics[trim=0cm 0cm 0cm 0.2cm, clip=true,width=0.6\textwidth]{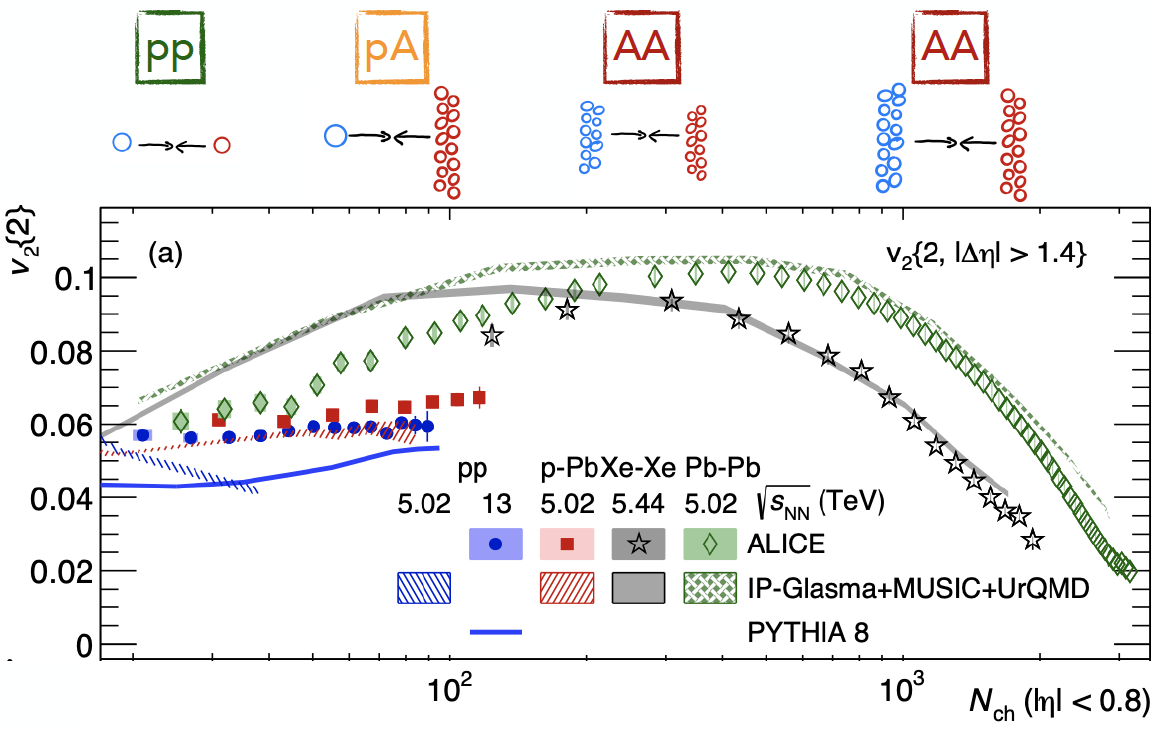}
\hspace{0.3cm}
\includegraphics[trim=0cm 0cm 0cm 0.2cm, clip=true,width=0.36\textwidth]{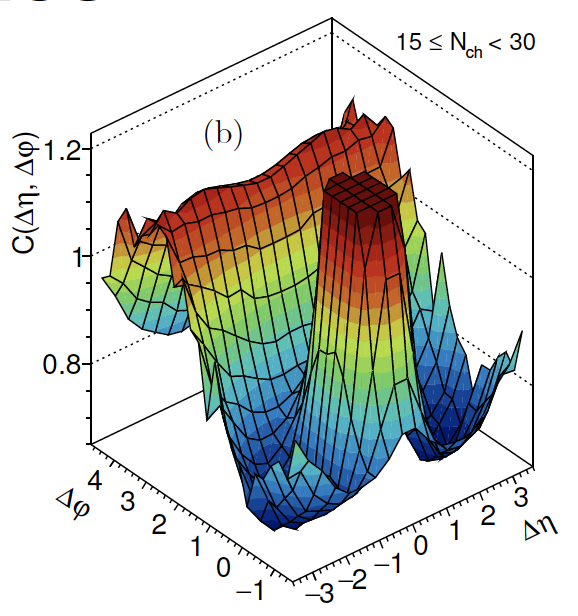}
 \caption{(left) Elliptic flow coefficient $v_{2}$ as a function of charged particle multiplicity in pp, p--Pb, Xe--Xe and Pb--Pb collisions~\cite{Acharya:2019vdf}. (right) Two-particle correlation $C(\Delta \eta, \Delta\varphi)$ for HM ep collisions~\cite{ZEUS:2019jya}.} \label{Fig::flowsystems}
\end{figure}

The $v_{2}$ coefficient is shown in Fig.~\ref{Fig::flowsystems} as a function of charged particle multiplicity in pp, p--Pb, Xe--Xe and Pb--Pb collisions. In the large collision systems, at large multiplicities, due to the initial geometry and strong interactions strong signs of collective effects are seen, which are well described by hydrodynamic calculations. In small collision systems (pp, p--Pb, peripheral Pb--Pb), the sizeable coefficients, which are similar in the different systems at the same multiplicity, could be indicative of collective effects. However, the data are neither described by hydrodynamic calculations nor by PYTHIA8 calculations, although the latter show a trend similar to data. Several detailed studies are ongoing to investigate the long-range correlations seen in the small systems and the continuous evolution of strangeness enhancement across collision systems~\cite{ALICE:2017jyt}. The ATLAS collaboration e.g.\ measured in pp collisions the elliptic flow as a function of charged particle multiplicity for inclusive and Z-tagged collisions (the impact parameter of the collision may be on average smaller)~\cite{Aaboud:2019mcw}, where neither a significant multiplicity dependence nor an influence due to the presence of the hard-scattering process is observed. In contrast to high-multiplicity (HM) pp collisions~\cite{Khachatryan:2015lva,Acharya:2019vdf}, no indication of collective behaviour is observed in elementary HM ep and e$^+$e$^-$ collisions as measured by the ZEUS~\cite{ZEUS:2019jya}, shown in Fig.~\ref{Fig::flowsystems}, and BELLE collaborations~\cite{Abdesselam:2020snb}. So far no signs of energy loss have been seen in small collision systems (see below). For further investigations, future measurements with large data samples and very HM pp collisions~\cite{ALICE-PUBLIC-2020-005} and collisions of light ions~\cite{Huss:2020whe} are foreseen.

\vspace{-0.2cm}
\section{Hard Probes (parton energy loss, jets, quarkonia)}
\vspace{-0.3cm}

\begin{figure}[htbp]
\includegraphics[width=0.5\textwidth]{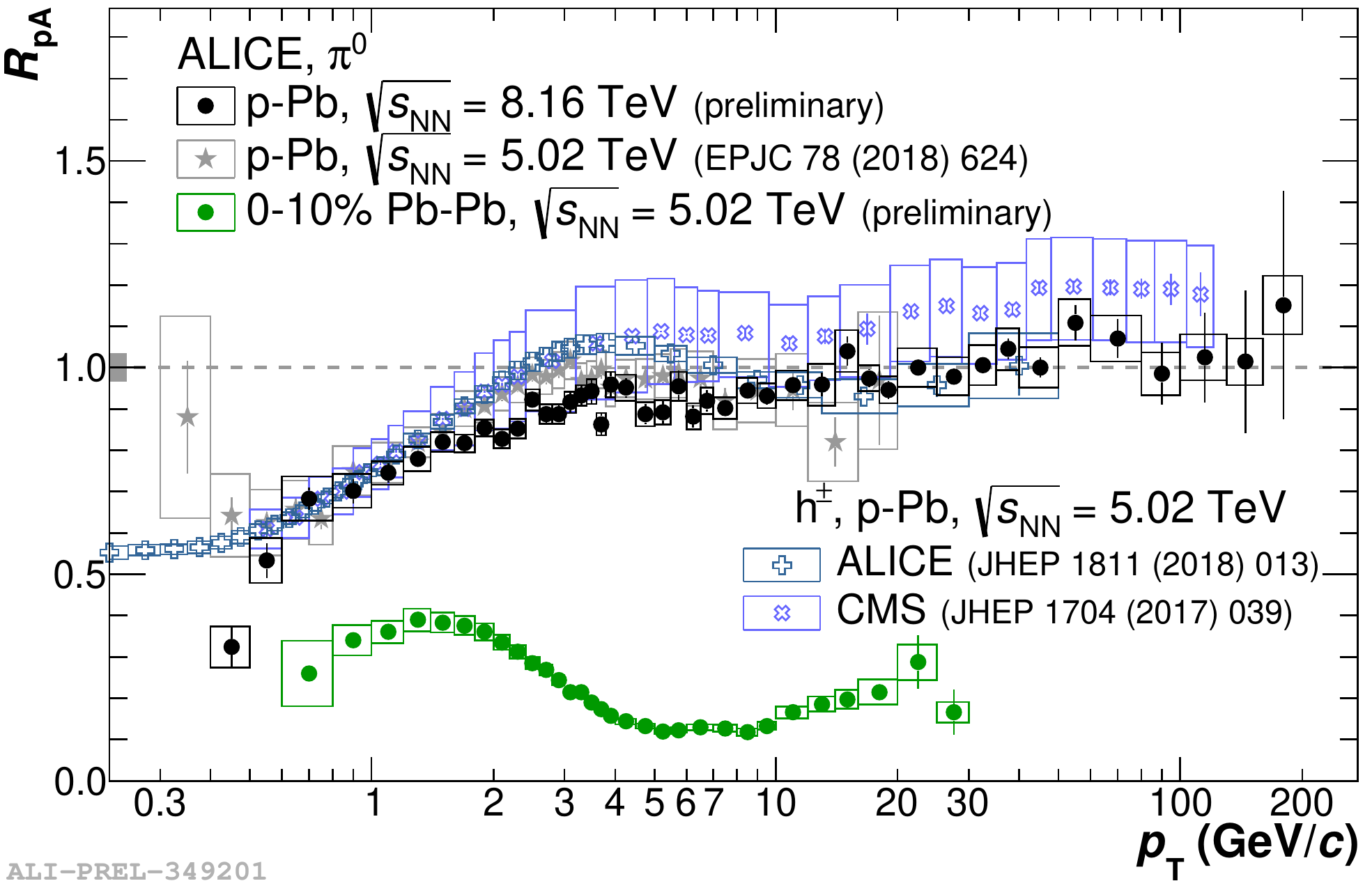}
\hspace{0.4cm}
\includegraphics[width=0.43\textwidth]{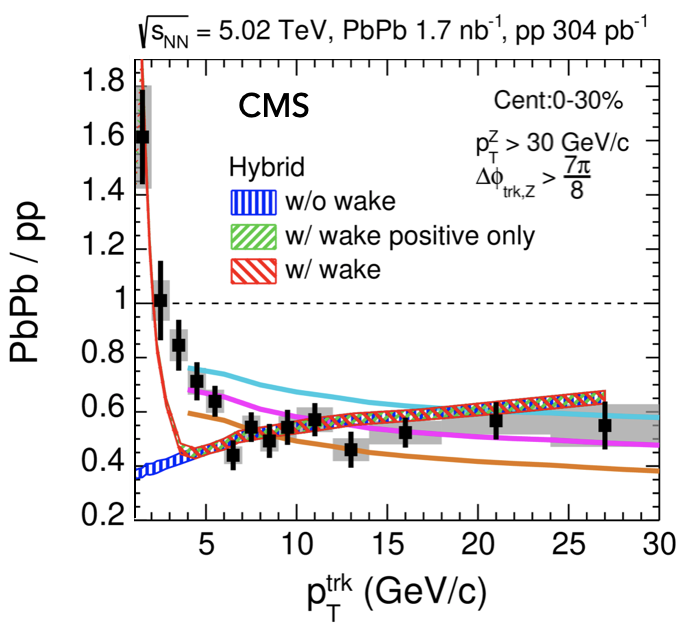}
\caption{(left) Nuclear modification factor of $\pi^{0}$ mesons and charged hadrons  in p--Pb and Pb--Pb collisions. (right) Ratio of Z boson-tagged charged hadron spectra in Pb--Pb and pp collisions~\cite{CMS:2020rxf}. } \label{Fig::RAA}
\end{figure}

In the large collision systems, such as in central Pb--Pb collisions, the yield of charged hadrons (see Fig.~\ref{Fig::RAA}) and of jets with large transverse momenta up to $p_{\rm T}~=$~900 GeV/$c$, as measured by the ATLAS collaboration~\cite{Aaboud:2018twu}, is strongly suppressed compared to pp collisions scaled by the number of binary collisions. The comparison with results from p--Pb collisions, where this effect is absent as seen in Fig.~\ref{Fig::RAA}, and theoretical calculations show that the suppression in Pb--Pb collisions is due to a final state effect, where the parton traversing the medium looses energy due to radiative and collisional energy loss. The large data samples at the LHC allow for many differential studies. For example, the CMS collaboration measured for the first time Z-tagged charged particle yields in Pb--Pb collisions, which are, as shown in Fig.~\ref{Fig::RAA}, compared to the expected yields in pp collisions strongly suppressed at high $p_{\rm T}$, but increase towards low $p_{\rm T}$~\cite{CMS:2020rxf}. The jet substructure was studied by the ALICE collaboration~\cite{ALICE:pubnotejets} as a function of the jet resolution parameter in Pb--Pb and pp collision using the grooming technique, where the large angle soft gluon radiation is removed to identify a hard splitting within the jet. The fully corrected data show that the jet substructure is modified by the hot medium, the selected hard splitting being narrower, which is described by e.g.\ models that include incoherent energy loss~\cite{ALICE:pubnotejets}.

\begin{figure}[tbp]
\includegraphics[width=0.43\textwidth]{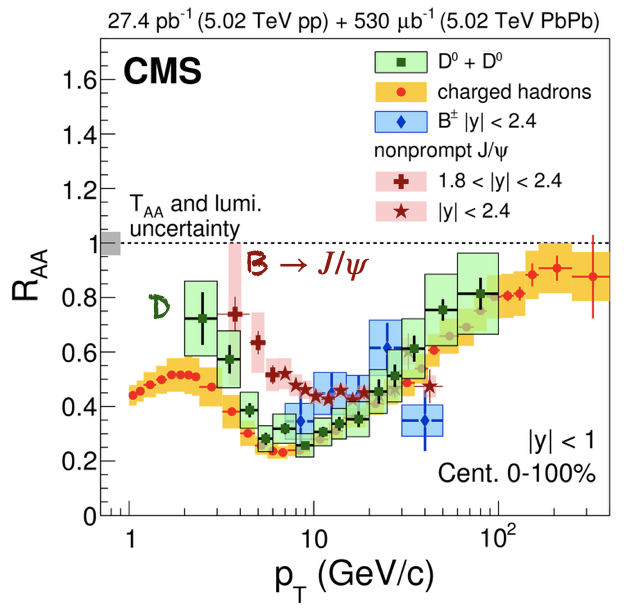}
\hspace{0.5cm}
\includegraphics[trim=0cm 0cm 1.0cm 1.0cm, clip=true,width=0.48\textwidth]{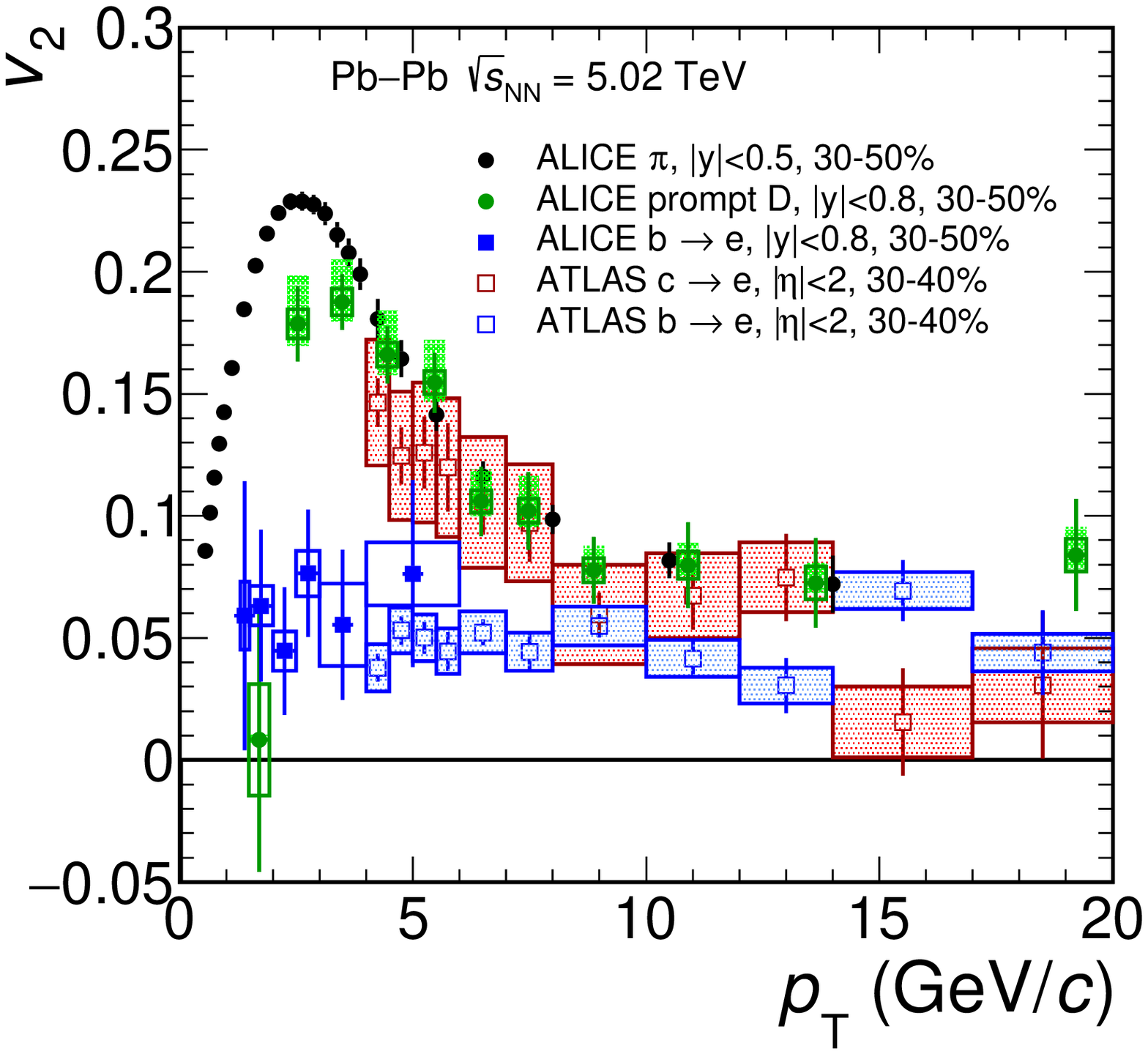}
\caption{Nuclear modification factor $R_{\rm AA}$ 
of charged hadrons (mainly pions), prompt D mesons, B$^{\pm}$, non-prompt J/$\psi$ as a function of $p_{\rm T}$ in Pb--Pb collisions at $\sqrt{s_{\mathrm{NN}}}=5.02$~TeV~\cite{Sirunyan:2017xss}. Elliptic flow $v_{\rm 2}$  of pions, prompt D mesons and electrons from charm- and beauty-hadron decays in the same collision system~\cite{Acharya:2020pnh,Acharya:2020qvg,Aad:2020grf}.\vspace{-0.5cm}
} \label{Fig::RAAandflow}
\end{figure}

Heavy quarks (charm and beauty) are created in initial hard scattering processes and thus experience the whole spatial and temporal evolution of a heavy-ion collision, thus providing essential information on the interactions of partons with the hot medium and its properties. Gluons are argued to loose more energy than quarks due to their stronger colour coupling to the medium. In addition, several mass-dependent effects (dead-cone effect and mass dependent spatial diffusion coefficient) are expected to influence the amount of energy loss of the heavy quarks. Thus a hierarchy of the parton energy loss is predicted, namely $\Delta E_{\rm gluon} > \Delta E_{\rm charm} > \Delta E_{\rm beauty}$, resulting in an expected ordering: $R_{\rm  AA}^{\pi} \le R_{\rm AA}^{\rm D} \le R_{\rm AA}^{\rm B}$ for pions (mostly originating from gluons), D, and B mesons. Measurements (e.g. those by the CMS collaboration in Fig.~\ref{Fig::RAAandflow}) show that $R_{\rm AA}^{\rm hadron} \approx R_{\rm AA}^{\rm D}$ for $p_{\rm T} \geq 5$ GeV/$c$. The $R_{\rm AA}$ is not only influenced by the energy loss, but also by the parton $p_{\rm T}$ spectrum and the fragmentation into hadrons. Theoretical calculations including these effects described the data well~\cite{Djordjevic:2013pba}. However, comparing with J/$\psi$ mesons from B-hadron decays and new measurements of prompt and non-prompt D mesons by the ALICE collaboration a quark mass depend energy loss at intermediate $p_{\rm T}$ is indeed observed. At high transverse momentum ($p_{\rm T}$ $\gtrsim$ 20~GeV/$c$), all hadrons including prompt J/$\psi$~\cite{Khachatryan:2016ypw} show within uncertainties surprisingly the same values of $R_{\rm AA}$. Differential new complimentary measurements by the ALICE and ATLAS collaborations (see Fig.~\ref{Fig::RAAandflow}) show that heavy quarks participate in the collective motion~\cite{Acharya:2020pnh,Acharya:2020qvg,Aad:2020grf}. Also here, at small $p_{\rm T}$, a mass-ordering effect becomes visible with electrons from beauty-hadron decays showing a positive $v_{2}$ with 3.8$\sigma$ significance indicating a partial thermalisation of beauty quarks in comparison to a large degree or complete thermalisation of charm quarks. At intermediate $p_{\rm T}$, the coefficients are similar for pions and D mesons, maybe due to a coalesence effect of a light and a charm quark. At high $p_{\rm T}$, the distributions merge suggesting that the path-length dependent energy loss dominates.

\begin{wrapfigure}{l}{0.48\textwidth}
\vspace{-0.4cm}
\includegraphics[width=0.5\textwidth]{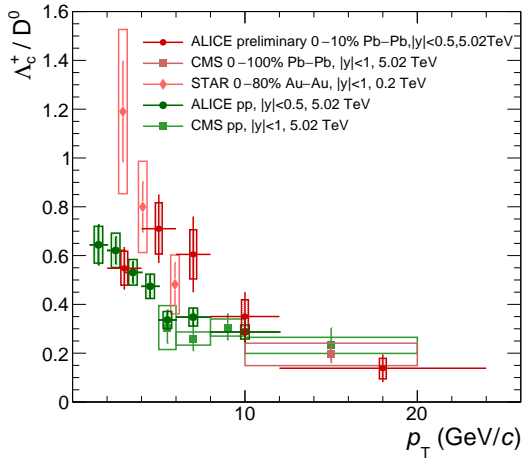}\vspace{-0.2cm}
\caption{$\Lambda_{\rm c}$/D$^0$ ratio as a function of $p_{\rm T}$ in pp and heavy-ion collisions~\cite{Acharya:2020uqi,Sirunyan:2019fnc,Adam:2019hpq}. } \label{Fig::Lambdac}
\vspace{-0.6cm}
\end{wrapfigure}

\begin{figure}[htbp]
\includegraphics[width=0.5\textwidth]{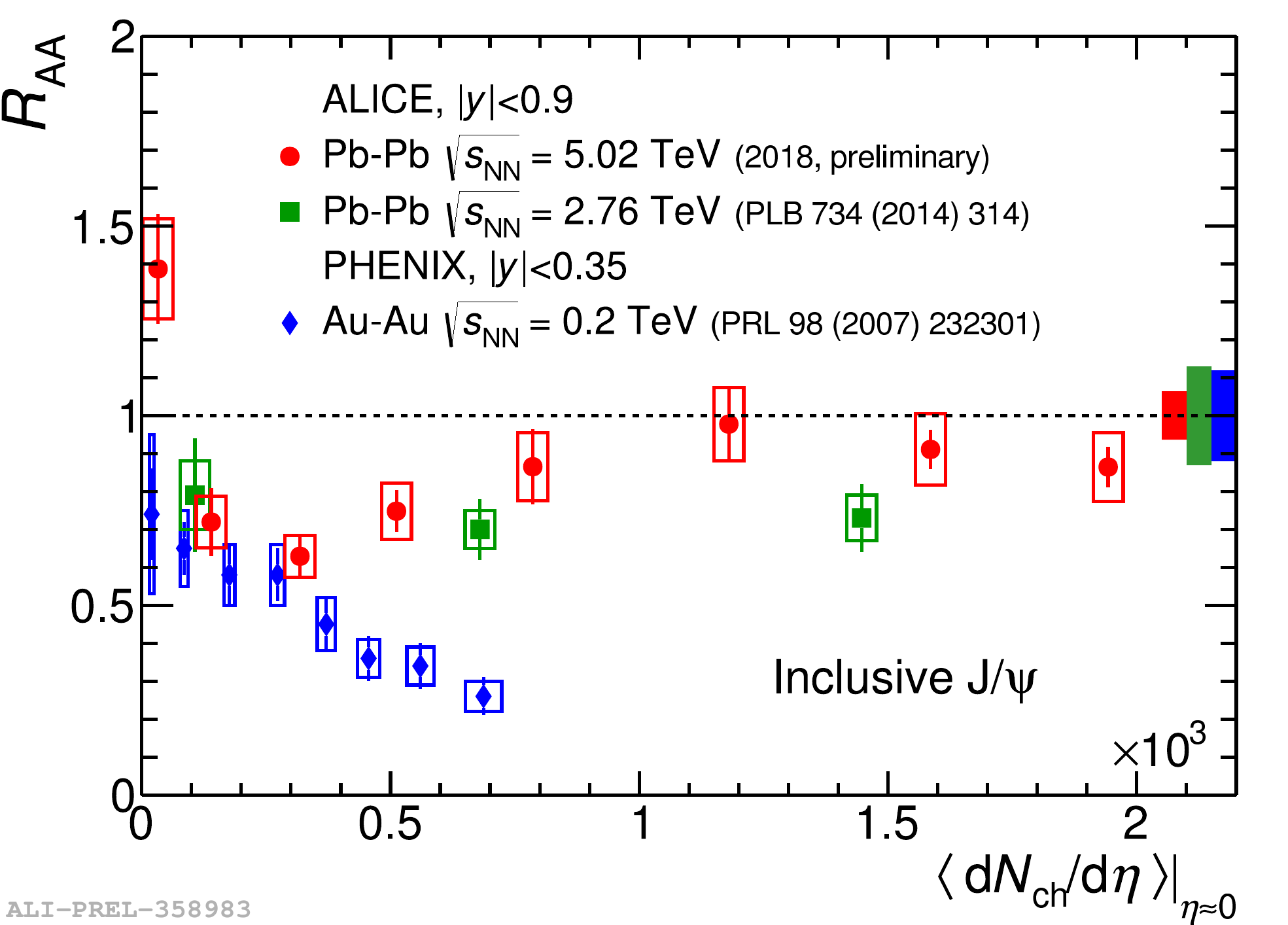}
\hspace{0.6cm}
\includegraphics[width=0.38\textwidth]{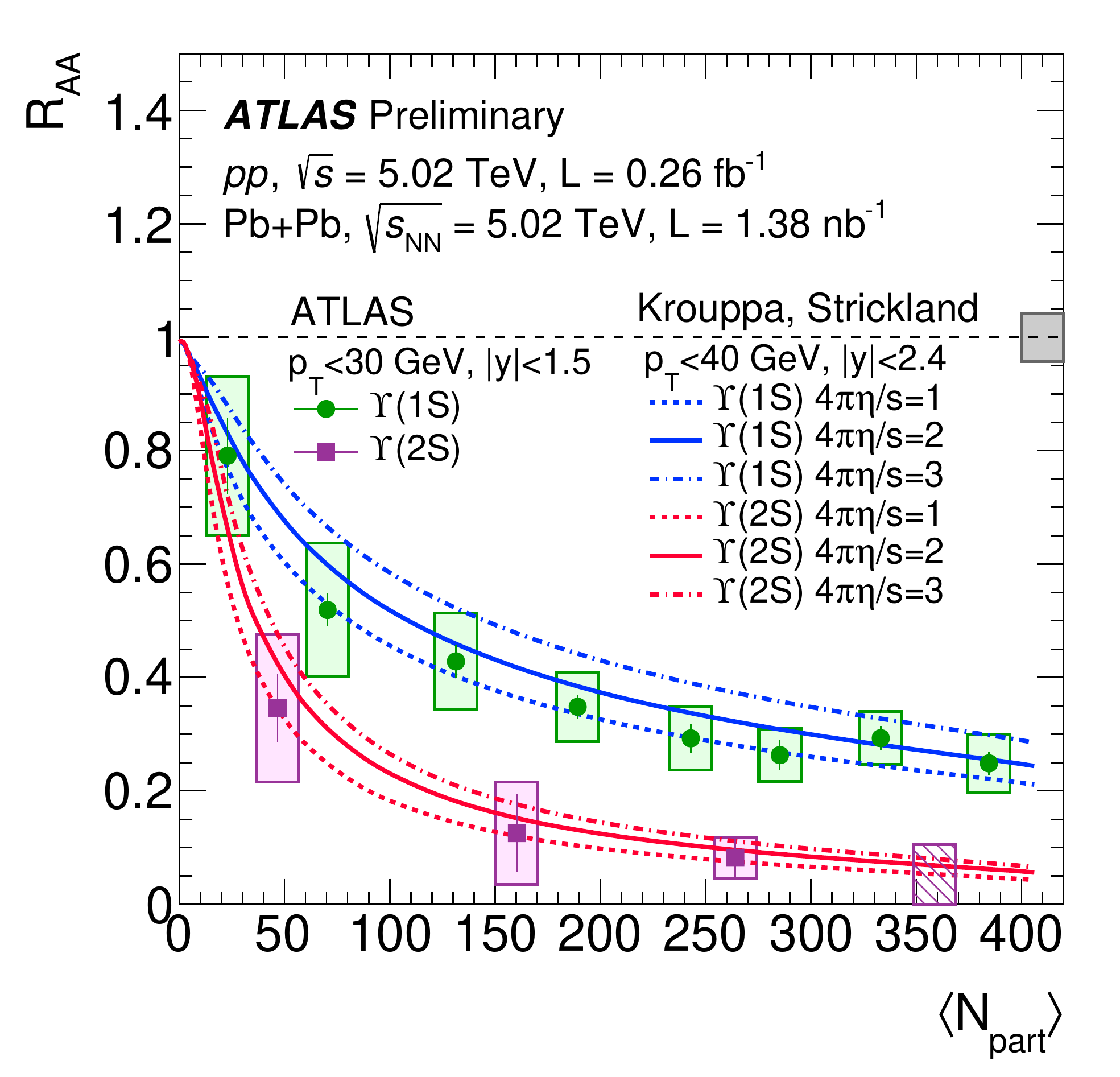}
\caption{$R_{\rm AA}$ of J/$\psi$ (left) and $\Upsilon$ states (right)~\cite{ATLAS-CONF-2019-054} as a function of charged-particle density and average number of participants, respectively.}\label{Fig::Quarkonia}
\end{figure}

To better understand the aforementioned mechanisms, yields of hadrons containing heavy quarks are also studied. The ALICE and the STAR collaboration measured the ratio of strange to non-strange D mesons in Pb--Pb collisions at $\sqrt{s_{\mathrm{NN}}}=5.02$~TeV and Au--Au collisions at $\sqrt{s_{\mathrm{NN}}}=0.2$~TeV. The ratios are higher by about a factor 2 in central collisions than the ones measured in pp collisions or PYTHIA calculations for the STAR result and well described by the statistical hadronisation model. The CMS collaboration finds the same behaviour in the beauty sector (B$_{\rm s}$/B$^+$). The data hint at an enhanced production of strange hadrons due to the strangeness-rich QGP. Thanks to the large available data samples, baryon-to-meson ratios can also be studied in the charm sector, which might provide insight into the hadronisation mechanism of charm baryons. The ratio of $\Lambda_{\rm c}$/D$^0$ in pp collisions shown in Fig.~\ref{Fig::Lambdac} is much larger compared to the one in e$^+$e$^-$ collisions ($\Lambda_{\rm c}$/D$^0$ $\approx 0.11$) and shows an even larger value for heavy-ion collisions as measured by the ALICE, CMS and STAR collaborations~\cite{Acharya:2020uqi,Sirunyan:2019fnc,Adam:2019hpq}. The measurements are qualitatively described by the statistical hadronisation model~\cite{Andronic:2017pug} and the Catania model with hadronisation via a combination of fragmentation and coalescence~\cite{Plumari:2017ntm}.\\
Quarkonia were proposed as a signature of the deconfinement in the QGP. It was predicted that quarkonium production would be suppressed due to a screening of the heavy-quark potential in the colour-deconfined medium~\cite{Matsui:1986dk}. With increasing temperature of the hot medium, the quarkonium states with decreasing radius were predicted to melt subsequently, thus providing a sort of `thermometer' of the QGP. Another competing mechanism is the production via (re)generation during the QGP phase or at hadronisation which depends  e.g.\ for the J/$\psi$ on the $c\overline{c}$ production cross section ($ N_{\rm{J}/\psi} \propto N^2_{c}$)~\cite{Grandchamp:2001pf,BraunMunzinger:2000px}. The LHC data allow the production mechanism of  J/$\psi$ to be elucidated. While the $R_{\rm AA}$ of J/$\psi$ decreases with increasing charged-particle density in Au--Au collisions at $\sqrt{s_{\mathrm{NN}}}=0.2$~TeV (see Fig.~\ref{Fig::Quarkonia}) the corresponding measurement at LHC energies shows at LHC full energy no suppression which is well in line with the regeneration scenario due to the large $c\overline{c}$ cross section at the LHC and is a clear signature of deconfinement. The differential studies vs $p_{\rm T}$ show an increase of the $R_{\rm AA}$ with decreasing $p_{\rm T}$, which is well described by theoretical transport models and the statistical hadronisation model that include the regeneration mechanism that dominates at low $p_{\rm T}$. The observed strong sign of collective effects of J/$\psi$ mesons~\cite{Acharya:2020jil}, including also the observed positive $v_{2}$ coefficient with a significance of 2.5$\sigma$ at midrapidity, confirms the regeneration scenario.\\
On the other hand, the suppression mechanism dominates for the bottomonium family as observed by an increasing suppression of $\Upsilon$(1S), $\Upsilon$(2S), and $\Upsilon$(3S) in measurements by the ALICE, ATLAS, CMS and STAR collaborations~\cite{ATLAS-CONF-2019-054,Sirunyan:2018nsz,Acharya:2020kls,Adamczyk:2016dzv}, see e.g.\ Fig.~\ref{Fig::Quarkonia}. The data are described by theoretical models that include the suppression mechanism and feedown from higher lying resonances.


\vspace{-0.2cm}
\section{Rare probes}
\begin{figure}[htbp]
\includegraphics[trim=0cm 0cm 0cm 0.9cm, clip=true,width=0.45\textwidth]{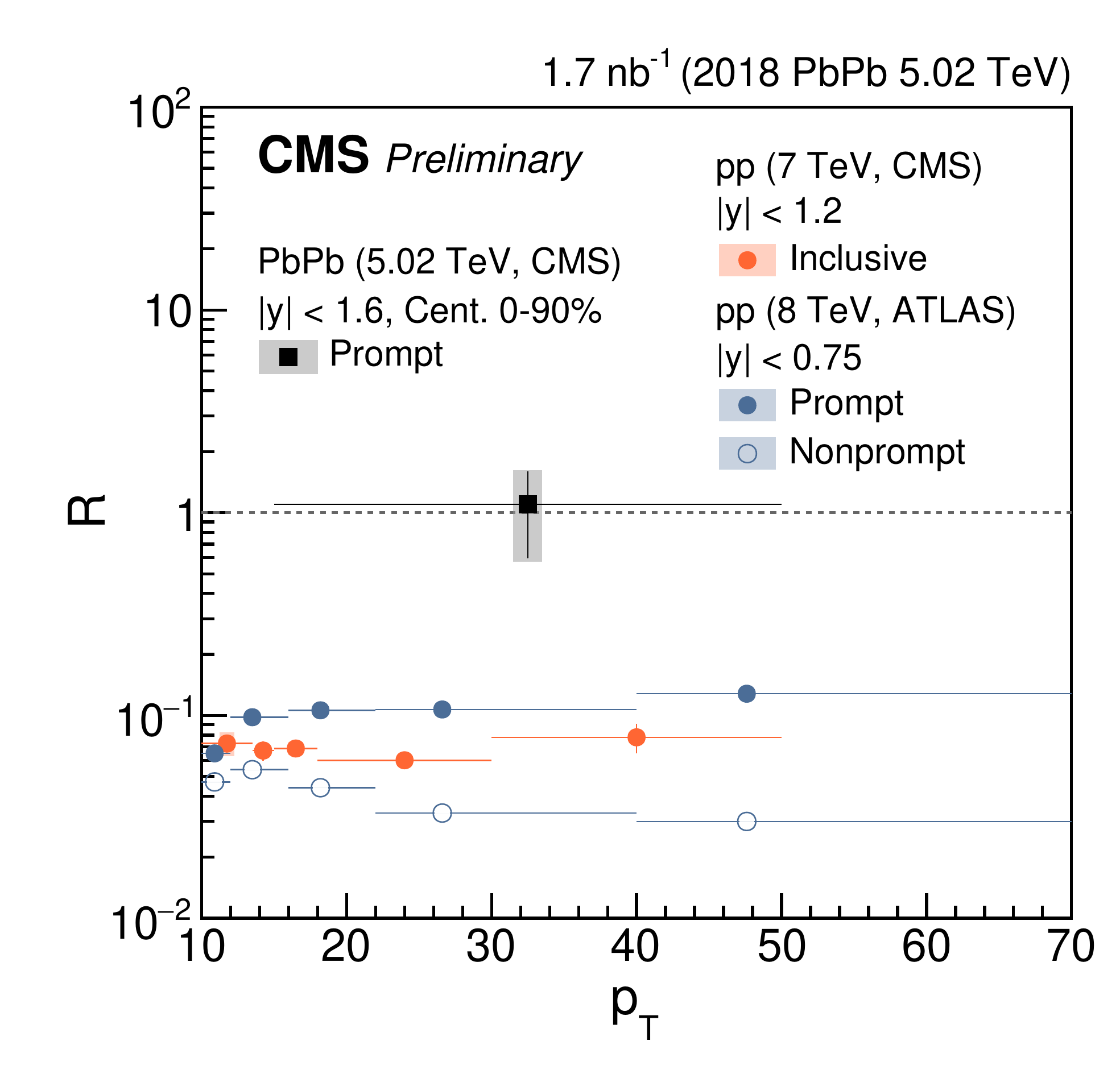}
\hspace{0.5cm}
\includegraphics[width=0.5\textwidth]{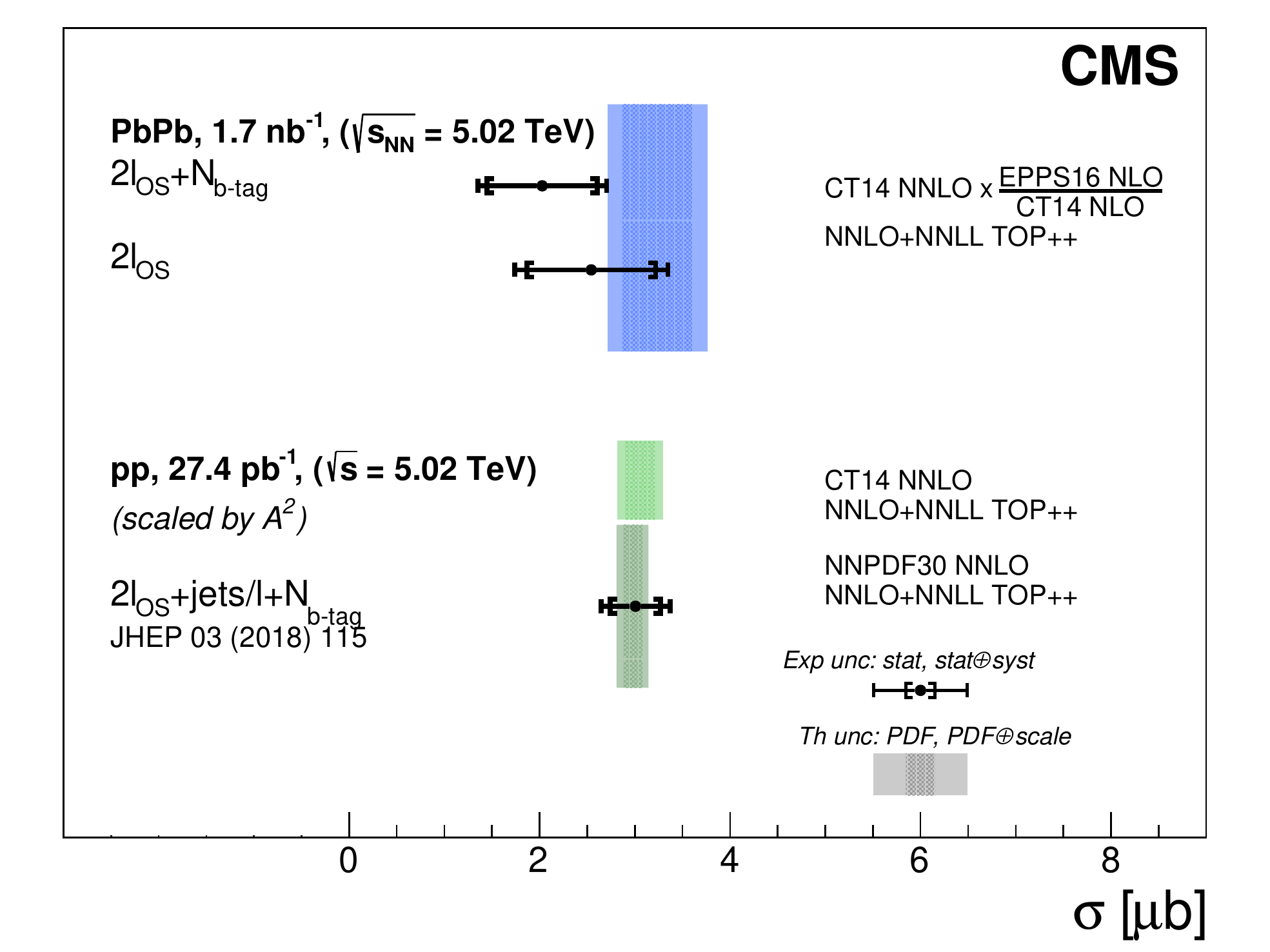}
\caption{(left) Ratio of the fully-corrected yields of $\chi_{\rm c1}$(3872) and $\psi$(2S) in pp and Pb--Pb collisions~\cite{CMS-PAS-HIN-19-005} and (right) t$\overline{\rm t}$ cross sections in pp and Pb--Pb collisions~\cite{CMS-PAS-HIN-19-001}.} \label{Fig::exotic}
\vspace{-0.5cm}
\end{figure}

\vspace{-0.3cm}
The large data samples recorded at the LHC, allow for new measurements of very rare probes. Studies of the $\chi_{\rm c1}$(3872) in heavy-ion collisions could provide additional insight into the production mechanism and the nature of this exotic particle. Figure~\ref{Fig::exotic} presents the ratio of the yield of $\chi_{\rm c1}$(3872) to $\psi$(2S) in pp and Pb--Pb collisions measured by the ATLAS and CMS collaborations. The ratio is larger in the heavy-ion collision, also because the CMS collaboration reported a significant suppression of $\psi$(2S) in Pb--Pb collisions. In future, it will be important to extend the measurement to low $p_{\rm T}$, where the regeneration mechanism dominates as seen for the J/$\psi$ meson (see above). \\
Evidence of top quark production in Pb--Pb collisions was for the first time reported by the CMS collaboration~\cite{CMS-PAS-HIN-19-001}. The measurement is performed via the semileptonic decay channel of W bosons with and without the presence of b-quark jets. The observed cross section $\sigma_{\rm t\overline{t}}$ as shown in Fig.~\ref{Fig::exotic} is compatible, but somewhat lower than pQCD calculations and the result from pp collisions correspondingly scaled. The measurement of top quark production in heavy-ion collisions might have the potential to probe the time structure of the QGP~\cite{Apolinario:2017sob} and thus will be an important study for the future, maybe for the LHC Run 5.


\vspace{-0.2cm}
\section{Impact beyond the physics of heavy-ion collisions}
\vspace{-0.3cm}
\begin{wrapfigure}{l}{0.5\textwidth}
\includegraphics[width=0.49\textwidth]{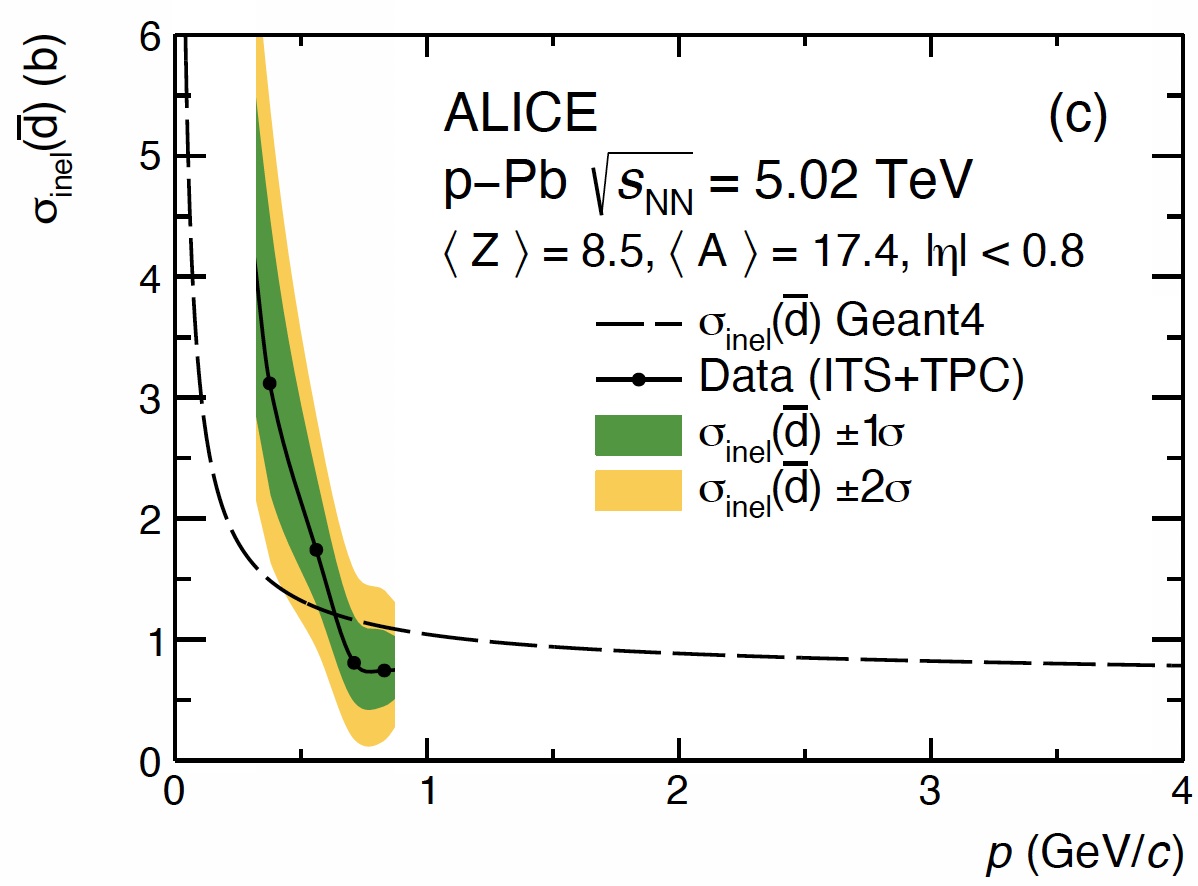}
\caption{Inelastic antideuteron cross section vs. momentum at which the interaction occurs~\cite{Acharya:2020cee}.}\label{figure:dcross}
\vspace{-0.6cm}
\end{wrapfigure}

Results from high-energy heavy-ion collisions also have impact on other fields of physics. Measurements of the cross sections of antinuclei provide important input for dark matter searches. The precise $p_{\rm T}$-differential antiproton cross section measurements in p--He collisions~\cite{Aaij:2018svt}, a measurement in fixed-target mode performed by the LHCb collaboration (He is injected into the beam pipe close to the interaction point), help to constrain theoretical models. The ALICE collaboration has measured for the first time the inelastic antideuteron--nucleus cross section at small momenta (see Fig.~\ref{figure:dcross}) using the detector as an absorber~\cite{Acharya:2020cee}. These kind of measurements will be extended to $^3{\overline{\rm He}}$ and $^4{\overline{\rm He}}$ in LHC Run~3 and 4.

\vspace{-0.4cm}
\section{Future of heavy-ion physics}
\vspace{-0.3cm}
The many new important results from the LHC and RHIC lead to an improved understanding of the interaction of partons with the hot medium, its properties and initial-state effects. They also constrain the nPDF global fits and provide input for other fields of physics. \\ 
With the ongoing upgrades (ALICE, ATLAS, CMS, LHCb) for LHC Run~3 and 4 an unprecedented level of precision will be reached and the search for rare probes significantly extended. Further, the European strategy for particle physics encourages the heavy-ion programme at CERN in the HL-LHC era. There is a plan for a next-generation LHC heavy-ion experiment: ALICE 3 constructed almost entirely from silicon~\cite{Adamova:2019vkf} for LHC Run~5 and beyond. In addition, projects such as sPHENIX and STAR at RHIC (incl.\ Beam Energy Scan), the fixed-target programme at CERN, facilities in the high net-baryon number density frontier (NICA, FAIR, ...) and the electron-ion collisions at EIC will provide further insight into different kinematic regimes of the QCD phase diagram.

\vspace{-0.4cm}
\begin{spacing}{0.9}
\setlength{\bibsep}{0.35\baselineskip}
\begin{footnotesize}

\bibliographystyle{JHEP}
\bibliography{bibliofile}
\end{footnotesize}
\end{spacing}

\end{document}

%% file: commands.tex
\newcommand{\twosevensix}  {$\sqrt{s}~=~2.76$~Te\kern-.1emV\xspace}
\newcommand{\five}         {$\sqrt{s}=5.02$~Te\kern-.1emV\xspace}
\newcommand{\twosevensixnn}{$\sqrt{s_{\mathrm{NN}}}=2.76$~Te\kern-.1emV\xspace}
\newcommand{\fivenn}       {$\sqrt{s_{\mathrm{NN}}}=5.02$~Te\kern-.1emV\xspace}